\documentclass{PoS}
\pdfoutput=1

\usepackage{amsmath}
\usepackage{graphicx}

\usepackage[sort&compress,numbers,merge]{natbib}
\usepackage{natbib,natbibspacing}

\newcommand{\preprint}{
  \begin{picture}(0,0)
    \put(0,80){{\rm\normalsize HU-EP-12/51}}
  \end{picture}}

\title{\preprint
Determination of $\boldsymbol{\Lambda^{\overline{\mathsf{MS}}}}$ from the
gluon and ghost propagators in Landau gauge}

\ShortTitle{Determination of $\Lambda^{\overline{\mathsf{MS}}}$ from the
gluon and ghost propagators in Landau gauge}

\author{\speaker{Andr\'e Sternbeck}\thanks{Supported by the EU commission (IRG
256594).}\\
  Institut f\"ur Theoretische Physik, Universit\"at
  Regensburg, D-93040 Regensburg, Germany\\
  E-mail: \email{andre.sternbeck@ur.de}}

\author{Kim~Maltman\\
        Department of Mathematics and Statistics, York Univ.,
        Toronto, ON, M3J 1P3, Canada\newline
        CSSM, School of Chemistry and Physics, University of
        Adelaide, SA 5005, Australia}

\author{Michael M{\"u}ller-Preussker\\
        Humboldt-Universit\"at zu Berlin, Institut f\"ur Physik,
  D-12489 Berlin, Germany}

\author{Lorenz von Smekal\\
        Institut f\"ur Kernphysik, Technische Universit\"at
  Darmstadt, D-64289 Darmstadt, Germany}

\abstract{We give an update on our lattice determination of
$r_0\LambdaMSb$ for different $N_f$. Our calculations employ the strong
coupling constant in the minimal MOM scheme for QCD in Landau gauge, and we
report here on our progress towards a quantitative understanding of the
intrinsic lattice discretization artifacts at large momenta. This is important
for a high-precision analysis, in particular for the unquenched calculations
for which the access to small lattice spacings is restricted by the available
gauge configurations.}

\FullConference{%
The 30 International Symposium on Lattice Field Theory -- Lattice 2012,\\
June 24-29, 2012\\
Cairns, Australia}

\newcommand{\Eq}[1]{{Eq.\,\eqref{#1}}}
\newcommand{\Fig}[1]{Fig.\,\ref{#1}}

\newcommand{\MM}{\mathsf{M\!M}}
\newcommand{\MSb}{\overline{\mathsf{M\!S}}}
\newcommand{\alphas}{\alpha_s}
\newcommand{\alphaMM}{\alphas^{\MM}}

\newcommand{\alphaMSb}{\alphas^{\MSb}}
\newcommand{\LambdaMSb}{\Lambda_{\MSb}}
\newcommand{\LambdaMM}{\Lambda_{\MM}}
\renewcommand{\Re}{\operatorname{\mathfrak{Re}}}      
\newcommand{\Tr}{\operatorname{Tr}}                   

\begin{document}

\begin{floatingfigure}[r] 
    \centering
    \parbox{7.7cm}{%
     \includegraphics[width=8.2cm]{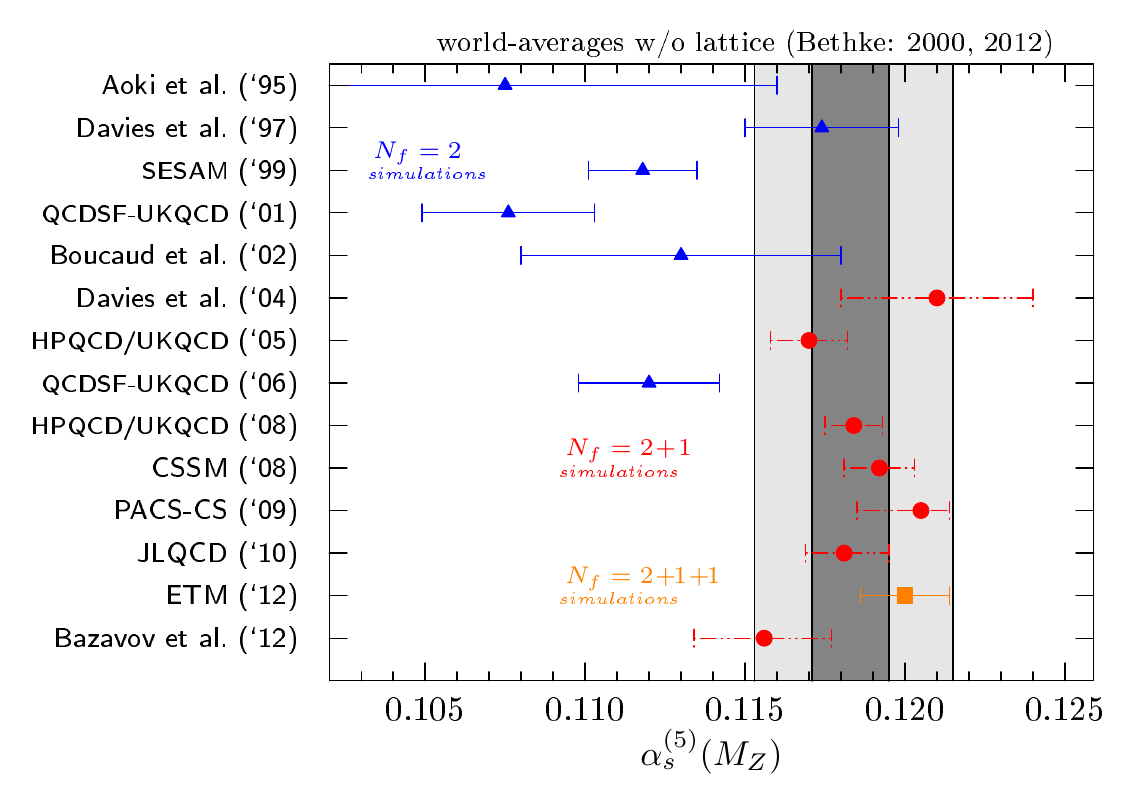}
     \caption{Lattice results for $\alphaMSb(M_Z$). Gray bars in the
      background mark the non-lattice averages from 2000 and 2012
      by Bethke (see, e.g., \cite{Bethke:2012jm})}}
     \label{fig:alphasMZ_lattice_results}
\end{floatingfigure}
\section{Introduction}

Lattice Monte Carlo calculations of the strong coupling constant
$\alphas=g^2/4\pi$ have been the subject of considerable interest in the 
lattice community (see figure). Over the years different schemes have been
devised and lattice calculations greatly improved, to the point that the recent
world average is dominated by lattice results \cite{Bethke:2012jm}.

Values for $\alphas$ are traditionally given at the Z-Boson mass in the $\MSb$
scheme for $N_f=5$ fermion flavors. Alternatively, one may quote, e.g.,
the $\MSb$ version of the dimensionful QCD $\Lambda$ parameter, 
$\LambdaMSb^{(N_f=5)}$. The latter is often adopted when quoting $N_f=0,2$
lattice results. While schemes other than $\MSb$ must be used on the lattice,
a 1-loop calculation suffices to yield $\LambdaMSb^{(N_f)}$ for different
$N_f$ in terms of the corresponding $\Lambda^{(N_f)}$ in these
other schemes.

In 2007 \cite{Sternbeck:2007br,*Sternbeck:2010xu} we introduced a
method for determining $\LambdaMSb^{(N_f)}$ using only Landau gauge
gluon and ghost propagators. 
These 2-point functions are easily accessible on the lattice and a product
of the gluon and ghost dressing functions, $Z_D$ and $Z_G$, defines the strong
coupling constant \cite{vonSmekal:1997is}
\begin{equation}
 \alpha^{\MM}_s(p) = \frac{g_0^2(a)}{4\pi}\ Z_D(p,a)\
Z_G^2(p,a)\qquad(\textrm{for}\;\;a\to0)
 \label{eq:alphaMM}
\end{equation}
whose running at large scales (momenta) $p$ is known up to 4-loop order
\cite{vonSmekal:2009ae}. In \Eq{eq:alphaMM}, $g^2_0$ is the
lattice coupling at lattice spacing $a$ and the suffix $\MM$ refers to the
Minimal MOM scheme, the underlying renormalization scheme of this coupling
\cite{vonSmekal:2009ae}. Note that $\alpha^{\MM}_s(p)$ is defined also beyond
Landau gauge. The first four $\beta$-function coefficients,
$\beta_0,\ldots,\beta_3$, can be found for general covariant gauges in
\cite{vonSmekal:2009ae}. Meanwhile this coupling has also been used by 
a French-Spanish collaboration
\cite{Boucaud:2008gn,*Blossier:2011tf,*Blossier:2010ky,*Blossier:2012ef} to
estimate $\LambdaMSb^{(N_f)}$ and a dimension-2 gluon condensate for different
$N_f$.

\section{Simulation details}

Our data for $\alphaMM$ is obtained on $N_f=0,2$ and $2\!+\!1$~ SU(3) gauge
field configurations. The quenched data is used to help quantify
systematic errors, primarily due to finite volume and lattice discretization
effects. $N_f=0$ configurations are thus generated (using the
standard Wilson gauge action) for different lattice spacings between
$a/r_0=0.186$ ($\beta=6.0$) and $a/r_0=0.037$ ($\beta=7.2$), and,
simultaneously, for different physical volumes $L^4$. 
We chose $L/r_0\approx2.5, 3.3$ and
4.3, and the $r_0/a$ values from \cite{Necco:2001xg,*Guagnelli:2002ia}.
With our lattice spacings we cover a large range of momenta, reaching well
into the perturbative regime (see below). This will allows us to fix
$r_0\LambdaMSb^{(0)}$ free of assumptions on the source of deviations from
pure 4-loop running at small momenta.

The unquenched configurations are kindly provided by the QCDSF collaboration.
For $N_f=2$ these are for the Wilson gauge action supplemented by
clover-improved Wilson fermions, while the $N_f=2\!+\!1$ configurations are for
a (tree-level) Symanzik improved gauge action and SLink fermions (see
\cite{Bietenholz:2011qq} for details).

All gauge configurations are fixed to lattice Landau gauge using an
iterative gauge-fixing algorithm. To guarantee high-precision the local
violation of transversality is not allowed to exceed \mbox{$\epsilon<10^{-10}$}
where, as usual,
$
\epsilon \equiv \max_x\, \Re\Tr\left[(\nabla_{\mu} A_{x\mu})(\nabla_{\mu}
    A_{x\mu})^{\dagger}\right]
$
and
\mbox{$ 
 A_{x\mu}\equiv
 \tfrac{1}{2iag_0 }(U_{x\mu}-U_{x\mu}^{\dagger})|_{\mathrm{tr.less}}
$}\,.

Standard techniques are also applied to calculate the gluon and ghost
propagators. Their dressing functions are extracted, as usual, by using the
exact tree-level structures of the respective lattice propagators. To reduce
lattice discretization artifacts further, we consider data only for diagonal
lattice momenta (strictest cylinder cut). Finally, data points for different
$\beta$ (and $\kappa$) are brought onto the common scale $r_0^2 p^2$, using the 
respective $r_0/a$ values. For $N_f=0$ we use the $r_0/a$ values from
\cite{Necco:2001xg,*Guagnelli:2002ia} and for $N_f=2$ the (new)
chirally-extrapolated values from QCDSF \cite{Bali:2012qs}.\footnote{Note that
the new $r_0/a$ values have changed, this explains why
$r_0\LambdaMSb^{(2)}$ is now larger than before
\cite{Sternbeck:2007br,*Sternbeck:2010xu}.} For $N_f=2\!+\!1$ we currently have
data only for $\beta=5.50$ for which $r_0/a\approx6.59$ \cite{BaliNajjar}.

\section{Quantifying the lattice corrections at large momenta}

Though the above standard tricks nicely smoothen the momentum dependence of our
$\alphaMM$ data, we cannot expect it to follow exactly 
the continuum 4-loop running, since, even away from lower momenta where
nonperturbative effects (and further down finite-volume effects) become
important, corrections due to the hypercubic lattice symmetry will
be needed. For a high-precision analysis it is essential to quantify these
corrections, which disappear only in the continuum limit. In particular
for our $N_f=2$ and 2+1 calculations a quantitative understanding of these
corrections is important. For them the access to sufficiently small
lattice spacings is still limited by the available configurations
($a/r_0\approx0.15\ldots0.12$), and (small but systematic) deviations from the
continuum 4-loop running at large momenta can be clearly seen in the data (see
Figs.\,\ref{fig:fitteddata} and \ref{fig:alphaMM_nf3} below).

A complete removal of hypercubic lattice artifacts could be achieved using the
so-called $H(4)$ method, as in 
\cite{Boucaud:2008gn,*Blossier:2011tf,*Blossier:2010ky,*Blossier:2012ef}. 
An alternative method, however, yields reduced statistical and systematic
uncertainties. To motivate this approach, consider
the gluon and ghost dressing functions, $Z_D$ and $Z_G$, in
(infinite-volume) lattice perturbation theory (henceforth ``LPT''). At 1-loop
order, e.g., these have the form
\begin{equation}
 Z_{D,G}(ap)=
  1+g_0^2\left[F_{D,G}(a^2p^2)+\Delta_{D,G}(ap)\right]\;
\stackrel{a\to 0}{=}\; 1+g_0^2 F_{D,G}(a^2p^2)
\end{equation}
with $F_{D,G}(a^2p^2)=\big(c^{D,G}_{11}\log(a^2p^2) +
c^{D,G}_{10}\big)$ the part that survives in the continuum limit.
The 1-loop coefficients $c^{D,G}_{10}$ and $c^{D,G}_{11}$ are known since
the eighties \cite{Kawai:1980ja}. For the hypercubic corrections, however, we
are interested in $\Delta_{D,G}(ap)$. These terms are non-zero for finite $a$,
depend on the momentum $ap$ and its direction \big(with
$ap_\mu=2\pi k_\mu/N_\mu$ and $k_\mu\in(-N_\mu/2,N_\mu/2]$\big) and represent
the 1-loop contributions to the hypercubic lattice artifacts. For lattice
hadron physics applications, e.g., it is common to calculate
such correction terms and subtract them from the lattice data for some
operators before determining the $\mathsf{RGI}$ scheme renormalization 
factors (see, e.g., \cite{Gockeler:2010yr}). Whether this removes
sufficiently the bulk of lattice artifacts depends on the quantity
being considered.

For the gluon and ghost propagators, to the best of our knowledge, nothing is
really known about $\Delta_{D,G}(ap)$. We therefore calculated them in LPT by
evaluating (mostly by numerical integration) the two Feynman diagrams (sunset
and tadpole) for the ghost self-energy and another seven for the gluon
self-energy, all for the whole range of external momentum ($ap$). In particular
the gluon self-energy diagrams containing 3- and 4-gluon vertices are quite
cumbersome.\footnote{A.S.\ thanks Holger Perlt for his help and support of (many
lines of) expressions for these vertices.} 

\begin{figure*}
 \mbox{\includegraphics[width=0.49\linewidth]{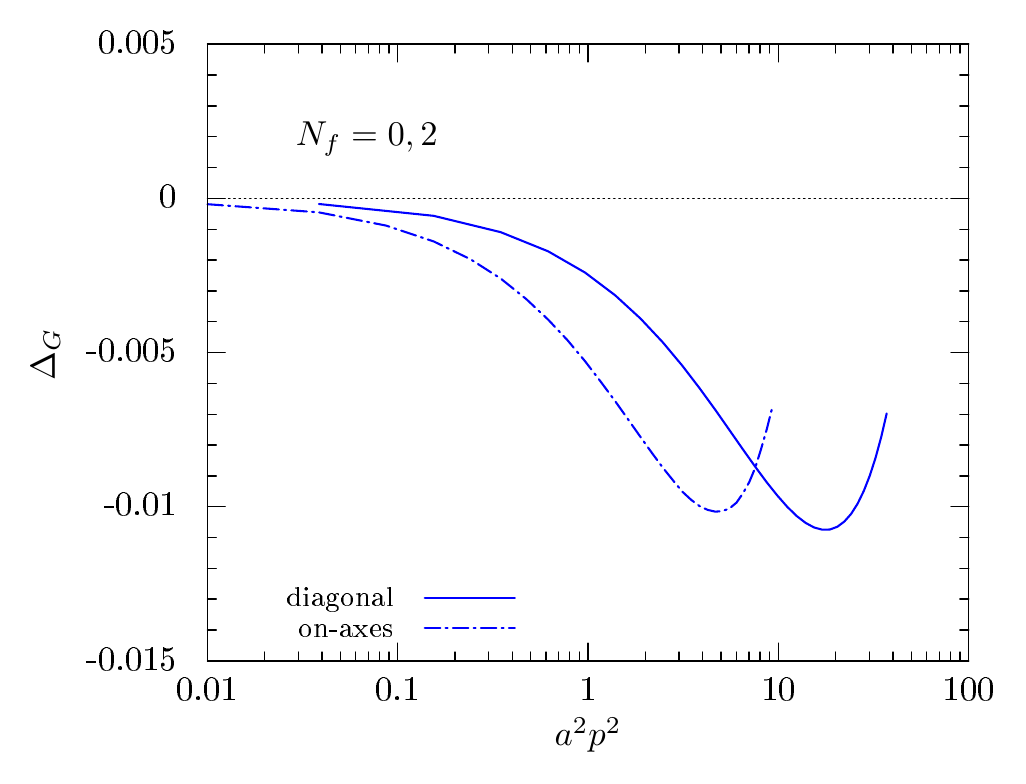}
 \includegraphics[width=0.49\linewidth]{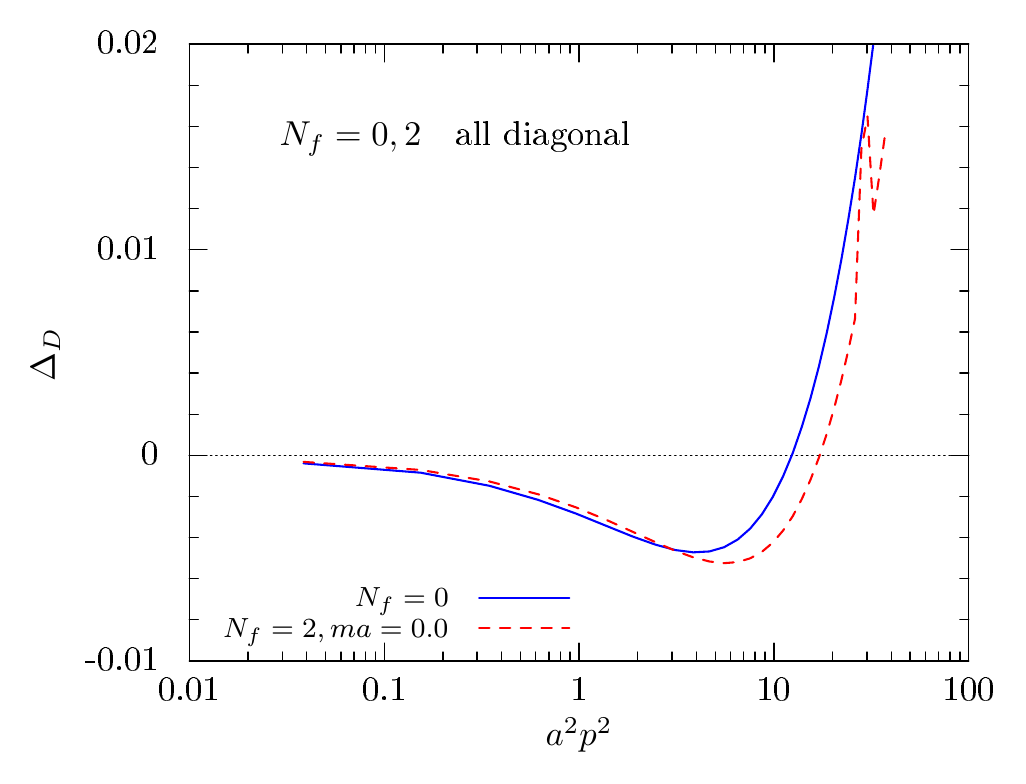}}
\caption{Hypercubic (1-loop) corrections, $\Delta_G$ (left) and
    $\Delta_D$ (right), versus $a^2p^2$. $\Delta_G$ is shown
    for on-axis (dash-dotted line) and diagonal momenta (solid line), while
    $\Delta_D$ only for diagonal momenta, but for $N_f=0$ (solid line) and
    $N_f=2$ (dashed line) massless fermions. Note that at 1-loop,
    $\Delta_G$ is independent of $N_f$.}
\label{fig:Delta_glgh}
\end{figure*} 

The outcome of our 1-loop LPT calculation is shown in \Fig{fig:Delta_glgh}.
The left panel shows $\Delta_{G}(ap)$ for two directions of $ap$,
diagonal and on-axis. As expected, the hypercubic corrections grow with $\vert
ap\vert$ and are smallest (but not at all zero) for diagonal $ap$. This is why
the widely used \emph{cylinder cuts} are so effective in removing a large
fraction of the lattice artifacts from the data, though these are not removed
completely. 
\begin{figure*}
 \centering
 \includegraphics[width=0.85\linewidth]{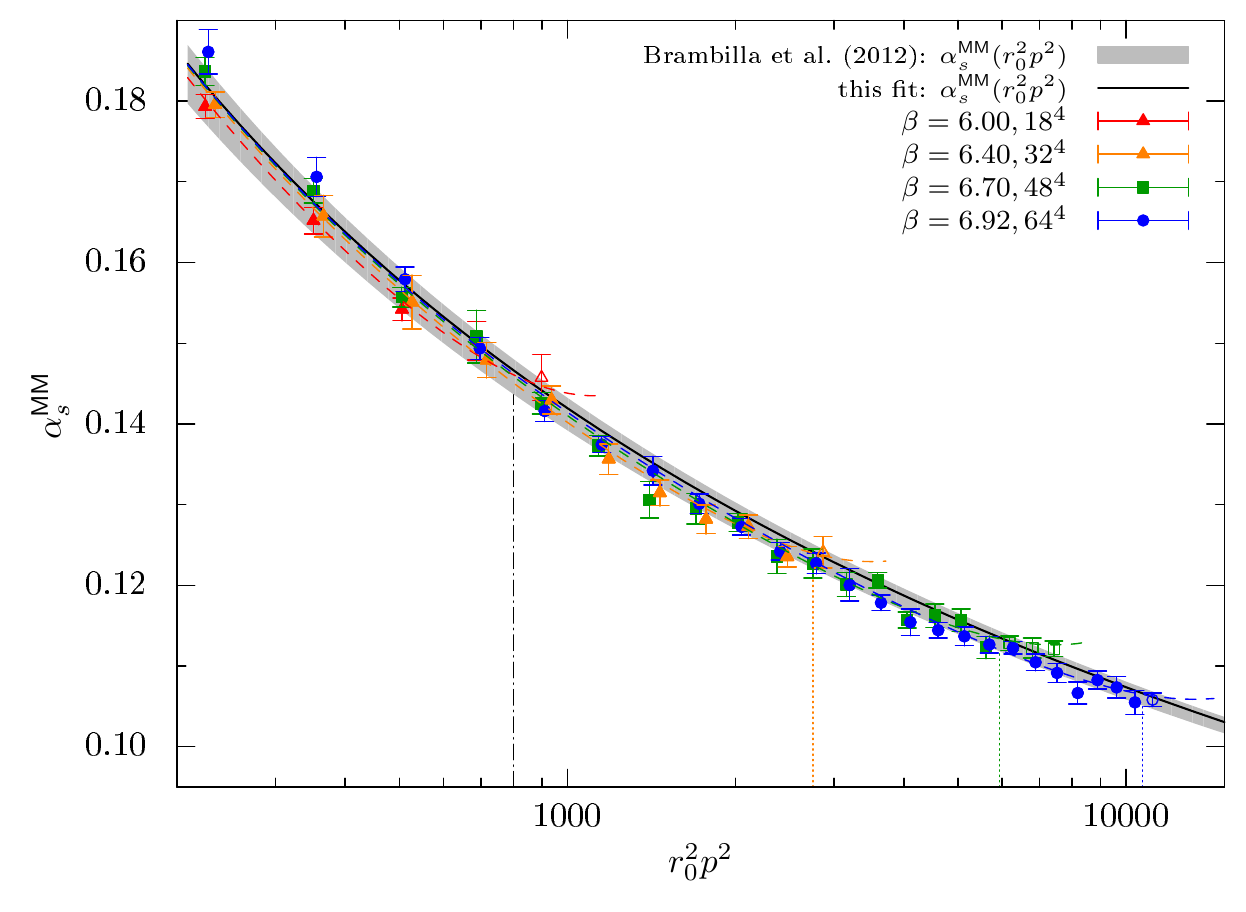}
 \includegraphics[width=0.85\linewidth]{%
 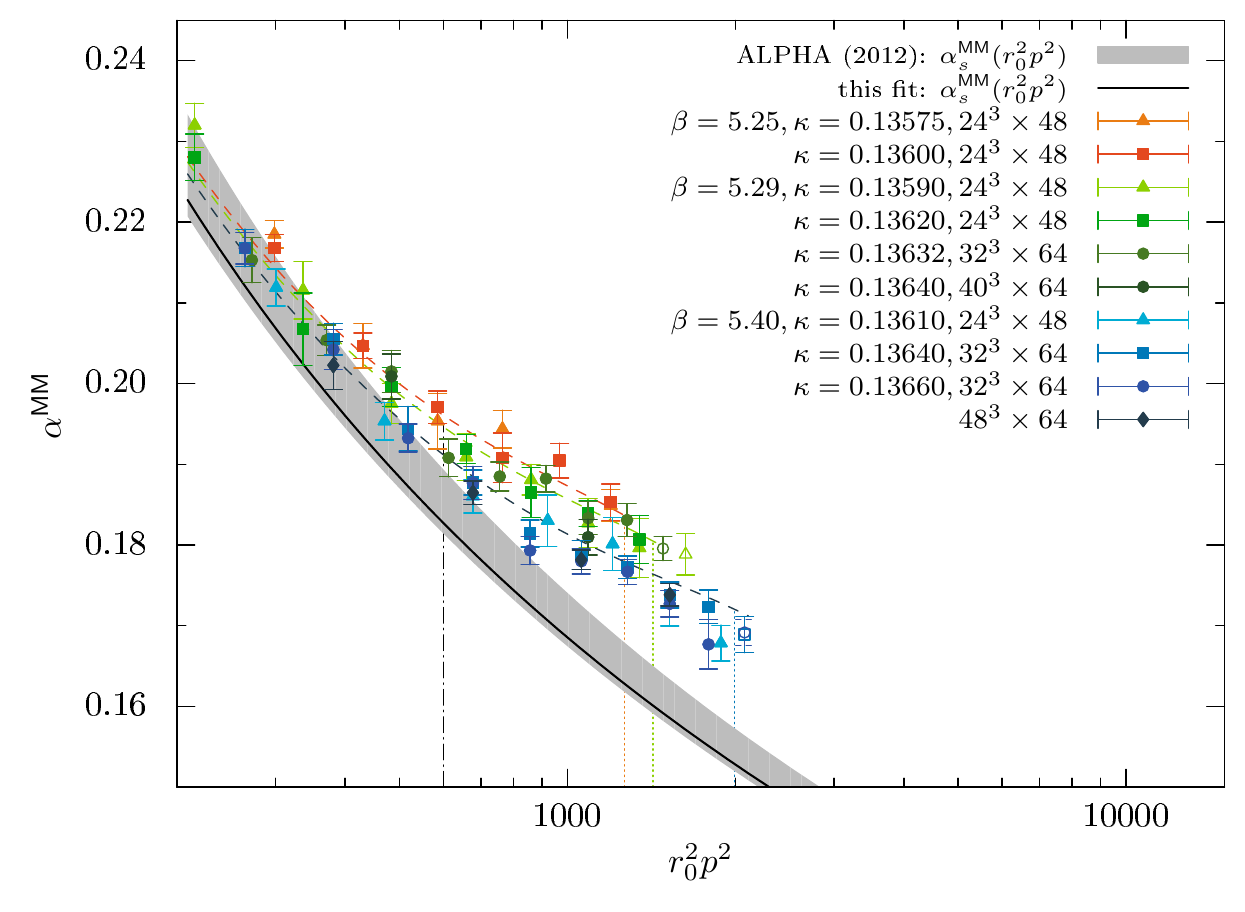}
\caption{$\alpha_{\mathrm{MC,sub}}^{\MM}$ versus $r_0^2p^2$ for $N_f=0$ (top)
  and $N_f=2$ (bottom). The gray band is $\alphaMM(r^2_0p^2)$ running at 4-loop
  using $r_0\LambdaMSb^{(0)}=0.637(32)$ and $r_0\LambdaMSb^{(2)}=0.789(52)$ as
  input \cite{Brambilla:2010pp,Fritzsch:2012wq}. The solid line (on this
  band) is $\alphaMM(r^2_0p^2)$ as it comes out from a fit of
  $\alpha_{\mathrm{MC,sub}}^{\MM}$ to \protect\Eq{eq:fitansatz} (no error
  band shown). For $N_f=2$ all data shown has been used (simultaneously) for
  the fit. For $N_f=0$ we show data for only one physical volume for simplicity,
  but seven more sets entered the fit. The (colored) dashed curves,
  connecting the data points, refer to $\alphaMM(r^2_0p^2)$ plus the
  fitted lattice corrections (see \protect\Eq{eq:fitansatz}). Vertical lines
  mark the fit window: a dashed-dotted line marks the lower bound
  $r^2_0p^2_{\min}=800$ \big($r^2_0p^2_{\min}=600$ for $N_f=2$\big) and dotted
  lines the upper bound $a^2p^2_{\max}=29$. Open symbols are for
  points where $a^2p^2>a^2p^2_{\max}$.}
\label{fig:fitteddata}
\end{figure*}
The right panel shows $\Delta_{D}(ap)$ for $N_f=0$ and $N_f=2$ massless
fermions. For simplicity only diagonal momentum results are
shown, the comparison of diagonal and on-axis momenta being similar to that seen
for $\Delta_{G}(ap)$. 

Looking at \Fig{fig:Delta_glgh} and the definition of $\alphaMM$ one sees 
that, to leading order, the lattice corrections to $\alphaMM(r_0p)$ are of
the form 
$
\alpha_s^{\MM}(r_0p) +g_0^4\left[\Delta_D(ap) +
 2\Delta_G(ap)\right] + O(g_0^6)\,.
$ 
Some of the corrections to the dressing functions will thus cancel in the
$\alphaMM$ data at larger $ap$. 

After subtracting $\Delta_{D}(ap)$ and $\Delta_{G}(ap)$ corrections from our
data for $Z_D$ and $Z_G$, we find, unfortunately, only minor reductions in the
size of lattice artifacts. Higher-loop corrections are thus needed, but
calculations beyond 1-loop LPT are not feasible because of their complexity.
Within the framework of Numerical Stochastic LPT this might be possible
\cite{DiRenzo:2009ni,*DiRenzo:2010cs,*Ilgenfritz:2010gu}. We have begun looking
into this.

In any case, from this exercise we know the remaining corrections should
be of the leading order $O(g_0^6)$. That is, a hypercubic Taylor expansion of
the (1-loop) subtracted $\alphaMM$ data for diagonal lattice
momenta, denoted $\alpha_{\mathrm{MC,sub}}^{\MM}$ henceforth, should
be of the form
\begin{equation}
\alpha_{\mathrm{MC,sub}}^{\MM}(r_0p,ap)
  = \alphaMM(r_0^2p^2) +
g_0^6\left[c_2\cdot(ap)^2 + c_4\cdot(ap)^4+ \ldots\right] + O(g_0^8)
 \label{eq:fitansatz}
\end{equation} 
where $c_2$ and $c_4$ are constants. In fact, this leading-order expression
describes the bulk of the lattice artifacts surprisingly well. We can fit our
lattice data, $\alpha_{\mathrm{MC,sub}}^{\MM}(r_0p,ap)$, \emph{simultaneously}
for different $\beta$ with the ansatz \Eq{eq:fitansatz} with only three
parameters: $c_2$, $c_4$ and $r_0\LambdaMM$ (resp.\ $r_0\LambdaMSb$), where the
latter parametrizes the 4-loop running of $\alphaMM$ \big(resp.\
$\alphaMSb$\big) in the continuum.

\section{Results}

To illustrate how well ansatz \eqref{eq:fitansatz} performs we show in
\Fig{fig:fitteddata} our (subtracted) $\alphaMM$ data versus $r_0^2p^2$ for
$N_f=0$ (top) and $N_f=2$ (bottom), together with the curves from a global fit.
A solid (black) line refers to $\alphaMM(r_0^2p^2)$, the strong coupling in the
continuum at 4-loop running, while the (colored) dashed lines refer to 
$\alphaMM(r_0^2p^2) + g_0^6\left[c_2\cdot(ap)^2 + c_4\cdot(ap)^4\right]$. Note
that $g_0^2=2N_c/\beta$ varies with $\beta$, but $c_2$ and $c_4$ are the
same for all data sets of same $N_f$. For the fit shown in \Fig{fig:fitteddata},
the fitting window starts at $r^2_0p^2=800$ ($r^2_0p^2=600$ for $N_f=2$) and
ends at $a^2p^2=29$. Other fit windows give similar results, but we see first
deviations from 4-loop running below $r^2_0p^2=500$. Due to the large lower
bound finite-volume effects play no role. For the purpose of this illustration,
we
therefore fitted the data for all available volumes, and also did not
distinguish the quark masses. This is suggested by the weak quark-mass
($\kappa$) dependence of our $N_f=2$ data (the corresponding pion masses
are $m_\pi\approx160\ldots720\,\text{MeV}$, see \cite{Bali:2012qs}). To not
overload \Fig{fig:fitteddata}, the $N_f=0$ data is shown only for one physical
volume.

For a comparison, we also show the continuum coupling $\alphaMM(r_0^2p^2)$
running at 4-loop (gray bands) using as input the recent values
$r_0\LambdaMSb^{(N_f=0)}=0.637(32)$ and $r_0\LambdaMSb^{(N_f=2)}=0.789(52)$ from
the literature \cite{Brambilla:2010pp,Fritzsch:2012wq}. Our
fits fully agree with these values.

\begin{floatingfigure}[r]
 \centering
 \parbox{7.7cm}{%
 \includegraphics[width=1.06\linewidth]{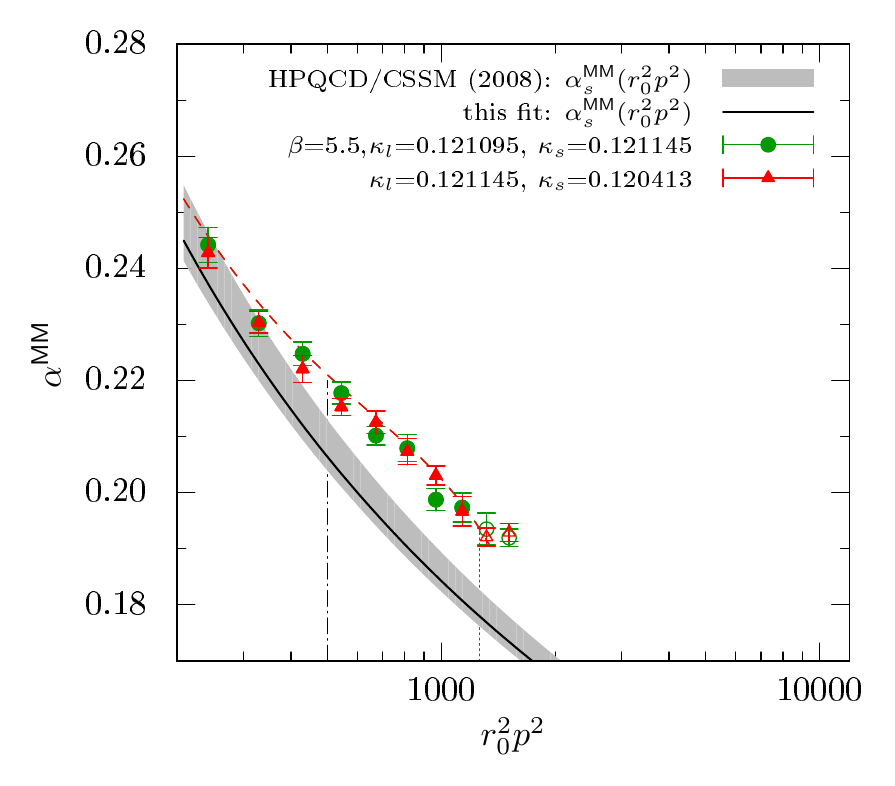}
 \caption{Two sets of our $N_f=2+1$ data and a first fit attempt similar to
 what was done for \protect\Fig{fig:fitteddata}.}%
}
 \label{fig:alphaMM_nf3}
\end{floatingfigure}
As mentioned above, there is also $\alphaMM$ data for $N_f=2\!+\!1$. We are
currently in the process of calculating $\alphaMM$ on all the different sets
provided by QCDSF to estimate $r_0\LambdaMSb^{(3)}$\,. So far we have data
only for one $\beta$ (see \Fig{fig:alphaMM_nf3}) on rather course lattices
($a/r_0\approx 0.15$). But a first fit, similar to those discussed above,
demonstrates already good agreement with the $\LambdaMSb$ from
\cite{Davies:2008sw,*Maltman:2008bx}. This is just a first
fit attempt, and a more thorough study will follow. But our new
approach of dealing with the lattice corrections (via \Eq{eq:fitansatz} and
restricting to data with diagonal momenta) seems to work nicely. It will allow
us to fix $r_0\LambdaMSb^{(N_f)}$, without the need to assume anything on the
nature of the deviations from pure 4-loop running at smaller momenta 
(higher-loop corrections, dim-2 condensate, etc.) as was necessary in
\cite{Boucaud:2008gn,*Blossier:2011tf,*Blossier:2010ky,*Blossier:2012ef}.

\vspace{-1.2ex}
\section{Summary}

We presented an update on our determination of $r_0\LambdaMSb^{(N_f)}$ for
different $N_f$. For this we use a renormalization scheme (Minimal MOM scheme
in Landau gauge) that only requires lattice data for the gluon and ghost
propagators in Landau gauge \cite{vonSmekal:2009ae}. In this contribution we
focused on a quantitative understanding of the hypercubic lattice
artifacts at large momenta. This is important for a high-precision analysis,
in particular for our unquenched calculations, for which the access to momenta
above 10~GeV is limited by the available gauge configurations. To make maximal
use of data at large lattice momenta ($a^2p^2\gg1$) we determined the leading
hypercubic lattice corrections to $\alpha_{\mathrm{MC,sub}}^{\MM}$.
This was achieved by a calculation of the contribution at 1-loop order in LPT
and secondly by a simultaneous fit of our lattice data (for diagonal momenta)
to \Eq{eq:fitansatz}. It then very well describes the leading lattice
corrections to $\alpha_{\mathrm{MC,sub}}^{\MM}$ with just three free parameters,
one of which is $r_0\LambdaMSb^{(N_f)}$. Values for $r_0\LambdaMSb^{(N_f)}$ for
$N_f=0$ and $N_f=2$ (and perhaps even $N_f=2\!+\!1$) will be given in a
forthcoming article. There we will give also more details on our data and fits,
and on the LPT calculations presented above.

{\bigskip\small\vspace{-1ex}
We are indebted to Holger Perlt for his help on getting the 1-loop LPT
calculations right. This work is supported by the European Union under the Grant
Agreements IRG 256594 and 249203. K.M.\ is supported by NSERC (Canada) and
L.v.S.\ by the Helmholtz International Center for FAIR. We thank the HLRN
(Germany) for the generous support of computing time over the years. 
}

\vspace{-2ex}
%

\end{document}